# Cryptanalysis of a Classical chaos-based cryptosystem with some quantum cryptography features

David Arroyo*
*Depto. de Ingeniería Informática, Universidad Autónoma de Madrid,
28049 Madrid, Spain
david.arroyo@uam.es*

Fernando Hernandez
*Escuela Técnica Superior de Ingenieros de Telecomunicación, Universidad Politécnica de Madrid
28040 Madrid, Spain*

Amalia B. Orúe
*Ortocrip S.L.
28290 Las Rozas (Madrid), Spain*

The application of synchronization theory to build up new cryptosystems has been a hot topic during the last two decades. In this paper we analyze a recent proposal in this field. We pinpoint the main limitations of the software implementation of chaos-based systems designed on the grounds of synchronization theory. In addition, we show that the cryptosystem under evaluation possesses serious security problems that imply a clear reduction of the key space.

## 1. Introduction

The connection between the basics of information encryption and the theory of dynamical systems is very well known from the seminal work of Claude Shannon [Shannon, 1949]. This connection has been profusely exploited from Baptista's work in 1998 [Baptista, 1998] and it has originated the so-called chaos-based cryptography. Although it is possible to build up secure chaos-based cryptosystems, along these years plenty of works have been published highlighting security and efficiency weaknesses of those encryption systems [Alvarez et al., 2011]. From a general point of view, chaos-based cryptosystems can be divided into schemes based on chaotic synchronization and those working in discrete time domain. Regarding the former group, their nature makes them vulnerable according to the security standards in cryptography. Certainly, it is not difficult to find works (see [Alvarez et al., 2011] for a survey of them) underlining that the properties of chaotic synchronization can be either applied to conduct encryption and to infer the secret key or part of the secret key of these type of chaotic cryptosystems [Orue et al., 2009, 2010].

In [Vidal et al., 2012] an encryption scheme was proposed on the grounds of chaotic synchronization. The authors of that paper sustain that their proposal incorporates some important characteristics of quantum communications. As a matter of fact, quantum communications can be used to ease key exchange through optical channels, but also in free-space [García-Martínez et al., 2013]. In this paper we show some limitations of what the authors of [Vidal et al., 2012] tipified as quantum properties of their encryption scheme. Moreover, along this paper we show that their cryptosystem suffers from some of the most relevant problems in analog chaos-based cryptography [1]. Namely, it is possible to get an estimation of some of the secret parameters of the cryptosystems just by direct observation of the information in the public communication channel. Additionally, the cryptosystem has some configuration problems that result in an efficency degradation.

---
*Corresponding author

The rest of the paper is organized as follows. First, the cryptosystem described in [Vidal et al., 2012] is introduced. In Sec. 3 some important shortcomings of the cryptosystem are discussed, whereas Sec. 4 is focused on the recovery of the secret key by an attacker. The main consequences of the previous perfomance and security analysis are provided in Sec. 5.

## 2. Description of the encryption technique

The encryption system defined in [Vidal et al., 2012] is determined by the following dynamical system:

$$\dot{x}_A = y_A + \varepsilon_x(x_B - x_A) \quad (1)$$
$$\dot{y}_A = \mu\, x_A + x_A \left(a(x_A^2 + z_A^2) + bz_A^2\right) \quad (2)$$
$$\dot{z}_A = w_A \quad (3)$$
$$\dot{w}_A = \mu\, z_A + z_A \left(a(x_A^2 + z_A^2) + bx_A^2\right) \quad (4)$$

$$\dot{x}_B = y_B \quad (5)$$
$$\dot{y}_B = \mu\, x_B + x_B \left(a(x_B^2 + z_B^2) + bz_B^2\right) \quad (6)$$
$$\dot{z}_B = w_B + \varepsilon_z(z_A - z_B) \quad (7)$$
$$\dot{w}_B = \mu\, z_B + z_B \left(a(x_B^2 + z_B^2) + bx_B^2\right) \quad (8)$$

where $\varepsilon_x \in [0.1, 1.1]$ and $\varepsilon_z \in [0.1, 1.1]$.

The above set of equations determines two identical hyperchaotic dynamical systems, controlled by three common parameters $a, b, \mu$; each one has four variables $x, y, z, w$. Both systems are interconnected through two coupling strength parameters $\varepsilon_x$ and $\varepsilon_z$, that help to achieve the synchronization of both systems. Once the systems are synchronized, there exists a common keystream that is applied for encryption. Consequently, the scheme introduced in [Vidal et al., 2012] is a stream cipher built upon chaotic synchronization. Accordingly, the communication protocol between Alice and Bob (the transmitter and the receiver) is secure if they share the parameters $a, b, \mu$ and choose values for $\varepsilon_x$, $\varepsilon_y$ and for the initial values leading to synchronization. More in detail, the encryption procedure comprises five stages:

**Stage 1.** Alice and Bob set up their respective dynamical systems using the same $a, b, \mu$ parameters; but with two different sets of random initial conditions of the variables and coupling strengths, that are kept secret by Alice and Bob and are not interchanged between them, nor published. The initial conditions of the variables $x_{A0}, y_{A0}, z_{A0}, w_{A0}$ and $x_{B0}, y_{B0}, z_{B0}, w_{B0}$ are generated at random in the range $[-0.5, 0.5]$, by Alice and Bob respectively. These initial conditions change randomly each time the communication protocol starts.

**Stage 2.** Alice and Bob are interconnected using a communication channel, through which the values of the variables $z_A$ from Alice and $x_B$ from Bob are interchanged. As the initial conditions of each system are different, the initial trajectories of the variables of each system will be different.

**Stage 3.** When synchronization is achieved, we have $x_A = x_B$, $z_A = z_B$, and thus the coupling terms $\varepsilon_x(x_B - x_A)$ of Eq.(1) and $\varepsilon_z(z_A - z_B)$ of Eq.(7) (which can be interpreted as feedback signals) vanish. This being the case, Alice and Bob detect that the synchronization has occurred and stop transmitting $z_A$ and $x_B$. Hence, if an eavesdropper is connected to the communication channel after this moment, she will not obtain enough information to break the system.

---

[1] In the context of chaos-based cryptography, those cryptosystems built upon the synchronization of the underlying dynamical systems are coined as analog chaos-based cryptosytems [Li et al., 2007].

**Stage 4.** Each system keeps computing a numerically generated trajectory without any kind of information exchange during some time. After this transient time, Alice and Bob check again whether there is complete or generalized synchronization, that is done by checking whether the reconstructed attractor of either $x_B$ or $z_A$ has only two positive Lyapunov exponents. In this case, a symbolic sequence is derived from the chaotic orbit. In [Vidal *et al.*, 2012] this transformation is defined by locating the local minima of the chaotic orbit, and assigning a "0" ("1") to that value if it is a negative (positive) value. In order to avoid the dynamical reconstruction of the chaotic orbit, the symbolic sequence is sampled according to the Shannon's rate.

**Stage 5.** Information exchange is concealed by a Vernam cipher, using as *one-time pad* keystream the symbolic sequence created in the previous step. The Vernam cipher consists of combining the bits of the plaintext with the bits of the ciphering sequence by the Boolean exclusive-or (XOR) function.

## 3. Performance analysis

The tradeoff between usability and security is the crux of modern cryptography. Any encryption system must guarantee security and pave the way for end-users adoption. In this concern it is critical to propose encryption algorithms with low computational needs, otherwise the resulting cryptosystems would be discarded by any potential user. This commitment is not met by the proposal given in [Vidal *et al.*, 2012].

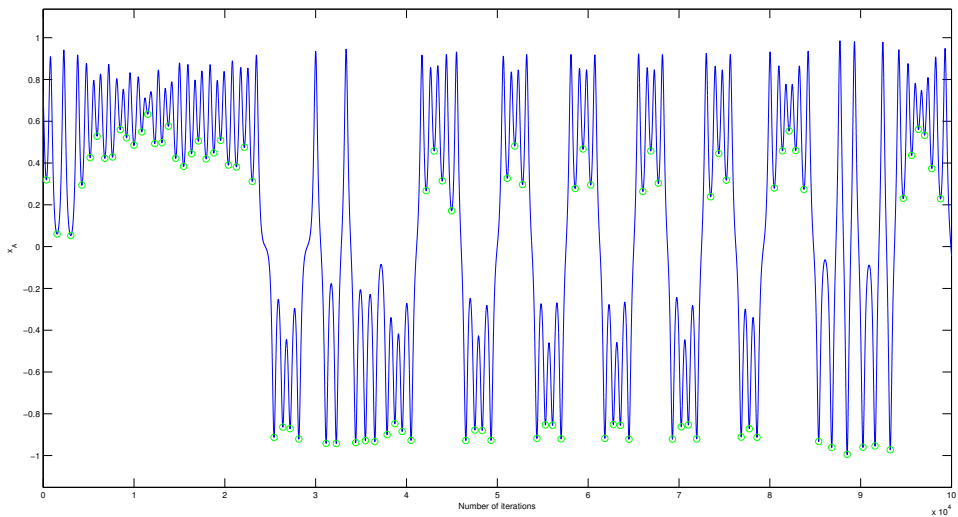

Fig. 1. Generation of symbolic sequences from a chaotic orbit according to the criterion defined in [Vidal *et al.*, 2012].

As it has been commented in Sec. 2, in the system under evaluation encryption is conducted by a keystream obtained through a quantization process focused on the local minima of the chaotic orbits determined by Eqs. (1) and (7). A major problem arises if encryption is performed on-line, since it is not possible to get a bit of the keystream as the plaintext is being processed. Therefore, the implementation of the cryptosystem as described in [Vidal *et al.*, 2012] calls for the buffering of plaintext until a new bit of the keystream is obtained.

Let us take as example the chaotic orbit in Fig. 1. That chaotic orbit contains $10^5$ samples, and the number of local minima is 92. Next, we define the throughput of the coding scheme as the ratio between the number of samples of a chaotic orbit and the number of local minima in the same orbit. Accordingly, we can conclude that the codification scheme proposed in [Vidal *et al.*, 2012] has a very low throughput[2].

---

[2]In addition, we have to take into account that in [Vidal *et al.*, 2012] it is further recommended to sample this original binary

In order to confirm this fact a set of 1000 chaotic orbits were generated using random values for the control parameters and the initial values. The number of samples per orbit was $10^5$, and the average throughput was 0.14%. Therefore, we can conclude that the coding technique proposed in [Vidal *et al.*, 2012] is far to be considered efficient and erodes the usability of the related cryptosystem. On this point it is advisable to recall that the maximum entropy of a dynamical system is obtained when orbit quantization is done through the so-called *generating partition* [Sinai, 1968]. Consequently, if our goal is to maximize the throughput of the coding procedure applied to chaotic orbits, we should consider some approximation of the generating partition without leaking information that an attacker could use to reconstruct the underlying dynamics [Arroyo *et al.*, 2009a].

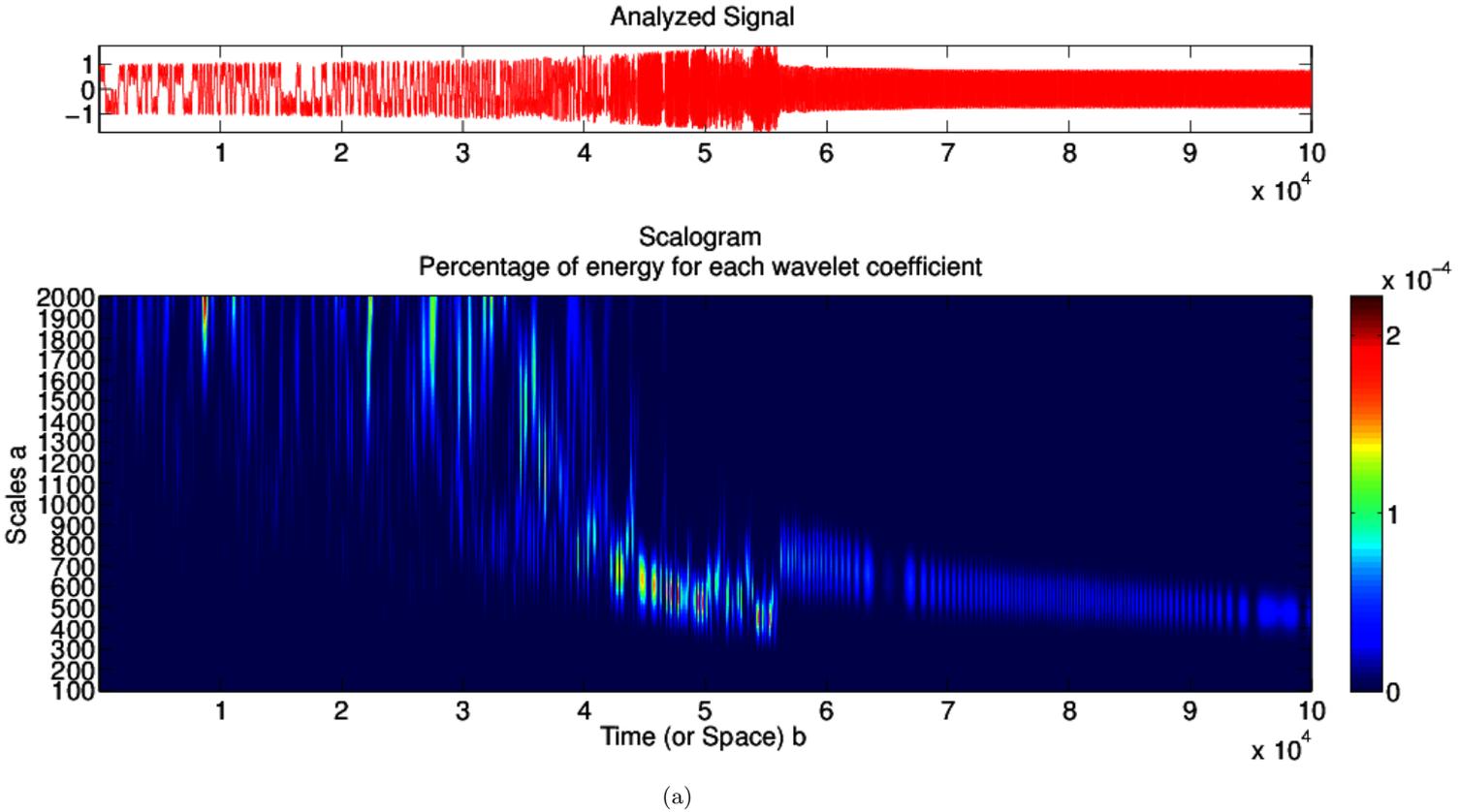

(a)

Fig. 2. Effect of finite precision computation on chaotic dynamics. In the top panel it is shown the orbit of length $10^6$ obtained from the system defined in [Vidal *et al.*, 2012] for $a = -0.924402423687748$, $b = 0.438971098170411$, $\mu = 0.711718876046661$, $x_{A_0} = 0.162590738289674$, $y_{A_0} = -0.442583550778422$, $z_{A_0} = 0.141686475255563$, $w_{A_0} = -0.194570102178438$, $x_{B_0} = 0.0601842136547941$, $y_{B_0} = 0.148286931043714$, $z_{B_0} = -0.307154096319608$, $w_{B_0} = 0.313998502860319$. The bottom panel gives an analysis of the dynamics of the orbits by means of a Morlet Continuous Wavelet Transform (CWT). The transition from chaos to limit cycle is given by the replacement of spread frequency components by more focused components in the scalogram of the CWT.

The application of continuous-time dynamical system to define discrete-time operations should be discarded, since it loads efficiency and reduce the throughput. As it was highlighted in [Arroyo *et al.*, 2009b], it is a much better option to select discrete-time dynamical systems to design digital cryptosystems. Moreover, the implementation of continuous-time dynamical systems is built upon numerical integration methods as Runge-Kutta's. These methods are parameter-dependent, which implies that the orbits calculated using

---

sequence to get a more robust protection against dynamics reconstruction by potential attackers. In specific, the authors of [Vidal *et al.*, 2012] establish as a good protection level to select one bit of every ten generated from the local minima of the chaotic orbits.

them are different for each selection of the set of parameters. As the Kerckhoffs' principle calls [Menezes *et al.*, 1997, p. 14], in the concrete field of chaotic cryptography it is necessary to include the parameters of the numerical integration methods as part of the public parameters or the secret key of the system.

Another key component in chaotic cryptography is the development of adequate procedures to avoid the so-called digital degradation [Wang *et al.*, 2016]. The implementation of chaotic dynamics in finite precision environments is not possible, since any chaotic orbit finally leads to a periodic orbit. As a matter of fact, the concretion of chaotic dynamics is not possible in finite precision unless some anti-control technique is incorporated [Hu *et al.*, 2014]. Along previous works this matter was underlined and typified as an intrinsic limitation of chaos-based cryptography [Arroyo *et al.*, 2011; Alvarez *et al.*, 2011], and in the case of [Vidal *et al.*, 2012] we have verified experimentally the erosion of chaos due to finite precision arithmetics[3]. Indeed, although two positive Lyapunov exponents for the system given in [Vidal *et al.*, 2012] is a necessary and sufficient condition to have chaos, from a practical point of view this is just a necessary condition. To back up this assertion we have conducted a series of experiments. In all the simulations performed we have verified that the initial chaotic behaviour of configurations with only two positive Lyapunov exponents collapses into a limit cycle, as it is depicted in Fig. 2. This transition from chaotic behaviour into a limit cycle has been detected using a time-frequency analysis [Chandre *et al.*, 2003; Chen *et al.*, 2011].

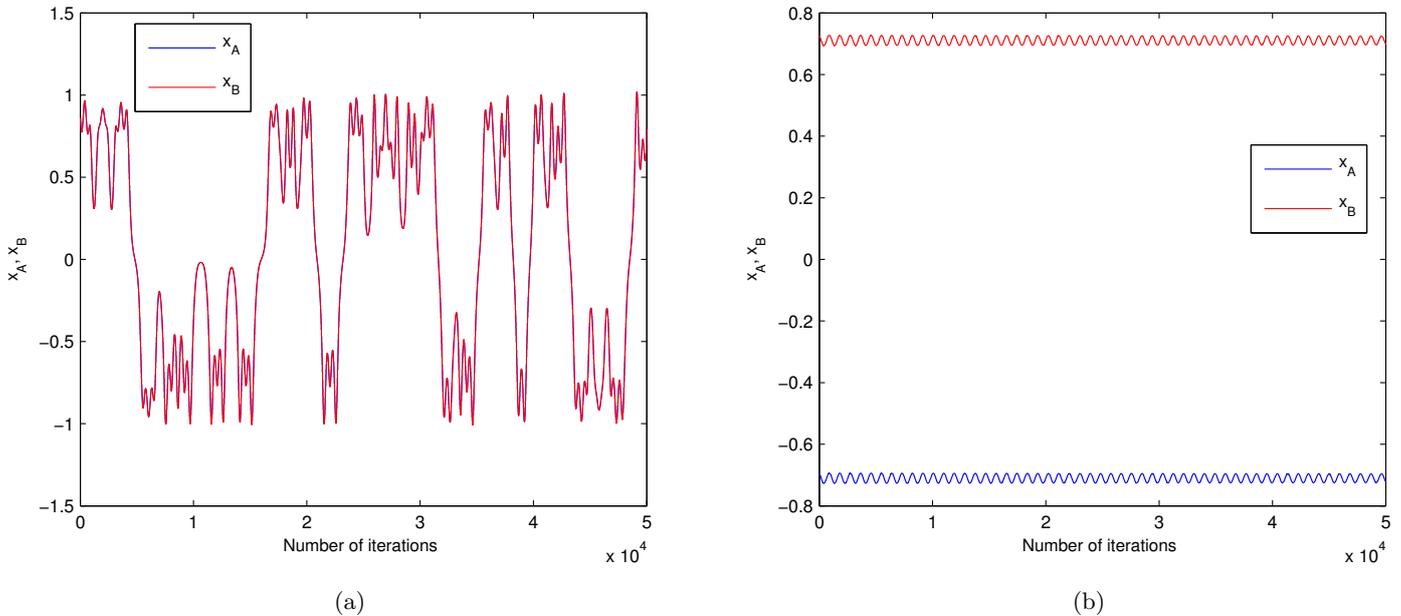

Fig. 3. Different behaviour of the dynamical system given in [Vidal *et al.*, 2012] for the same set of control parameters ($a = -0.924402423687748$, $b = 0.438971098170411$, $\mu = 0.711718876046661$) and different initial conditions (in (a) the initial conditions are $x_{A_0} = 0.162590738289674$, $y_{A_0} = -0.442583550778422$, $z_{A_0} = 0.141686475255563$, $w_{A_0} = -0.194570102178438$, $x_{B_0} = 0.0601842136547941$, $y_{B_0} = 0.148286931043714$, $z_{B_0} = -0.307154096319608$, $w_{B_0} = 0.313998502860319$; in (b) the initial conditions are $x_{A_0} = 0.162590738289674$, $y_{A_0} = -0.442583550778422$, $z_{A_0} = 0.141686475255563$, $w_{A_0} = -0.194570102178438$, $x_{B_0} = 0.473704902674984$, $y_{B_0} = 0.472305555688457$, $z_{B_0} = 0.143698049421405$, $w_{B_0} = 0.360098876854161$). The initial conditions of the transmitter are the same for both configurations: only the initial conditions of the receiver are modified.

---

[3]Furthermore, as it is shown in [Li, 2003], the adoption of floating-point over fixed-point computation implies additional problems. These problems are even more critical if we consider mobile devices as a possible application context (https://developer.android.com/training/articles/perf-tips.html#AvoidFloat, Last accessed 2016-08-31).

### 3.1. *Synchronization problems*

According to the authors of [Vidal *et al.*, 2012], their cryptosystem possesses quantum properties and can be used in VoIP communications. Regarding the quantum properties, it is possible to confirm that they are impaired by the inner characteristics and restrictions of chaotic synchronization in their setup. Taken for granted the analogy between chaotic synchronization and quantum communications, the coupling between the transmitter and the receiver should be of such a nature that synchronization is guaranteed when both ends of communication select initial conditions random and independently. In the case of the dynamical system selected in [Vidal *et al.*, 2012] this is not always satisfied. As the authors underline, the set determined by Eqs. (1) and (7) is not attracting. Futhermore, the existence of riddle basins in the synchronization manifold is both a protection against brute-force attacks and a problem in a practical context where the implementation of the dynamical systems must be done with finite precision computations. To highlight this last consideration we have carried out a rigorous analysis of the stability of the synchronization manifold. First, we have selected a set of control parameters and initial conditions determining only two positive Lyapunov exponents. Second, we have updated randomly the initial conditions of the receiver, whereas the rest of elements conforming the secret key and public parameters were kept as they were selected during the first stage of the experiments. From the experiments we have realized we got a rate of 33% of configurations where synchronization between the transmitter and the receiver is not achieved when the initial conditions of the receiver are changed. To further illustrate this matter, we show in Fig. 3 an example of how a synchronization state of the system given in [Vidal *et al.*, 2012] can be destroyed just by modifying the initial conditions of the receiver. This situation means that the initial conditions of the trasmitter and the receiver cannot be established independently, and this is against the quantum properties claimed by the authors of the cryptosystem in [Vidal *et al.*, 2012].

On the other hand, we should evaluate whether the synchronization between transmitter and receiver is affected by time-delays in the communication channel. Certainly, if one considers a setup as the one given in Fig. 4 and introduces a transmission delay, then synchronization should not be degraded. However, it is possible to confirm experimentally that a small time-delay could determine two different fixed points in the transmitter and receiver (see Fig. 5), which is not the expected chaotic sychronized state.

### 3.2. *Comments about the selection of adequate values for the control parameters and the secret key*

As it has been highlighted in recent cryptanalysis works [Liu *et al.*, 2015; Li *et al.*, 2013], the selection of adequate values of the control parameters and initial conditions is a critical point in chaotic cryptography. In this regard we have to underline that there is not an explicit definition of key space in [Vidal *et al.*, 2012], since there is not a clear description of the control parameters and initial conditions to force hyperchaotic regimen. Nevertheless, this paper is based on [Vidal, 2011], where two different configurations leading to hyperchaotic behavior are provided:

- $a = -1$, $b = 1.1$, $\mu = 0.88$ [Vidal, 2011, p.53].
- $a = -1$, $b = 0.9$, $\mu > 0$ [Vidal, 2011, p. 46].

Nonetheless, we have verified in our experiments that these two configurations not always determine a hyperchaotic behaviour. Moreover, hyperchaoticity can be achieved using other configurations from those given in [Vidal, 2011]. The problem here is that the criterion to select adequate values of the control parameters and the initial conditions is based on the evaluation of Lyapunov exponents, which resorts to the computation of millions of iterations of the differential equations [Vidal & Mancini, 2009, p. 722]. This fact determines a degradation of the efficiency of the cryptosystem described in [Vidal *et al.*, 2012].

## 4. Security analysis

### 4.1. *First considerations about the key space*

Recalling Kerkchoffs' principle, the definition of a cryptosystem must incorporate the clear concretion of its secret key, i.e., the cryptosystem's secret parameters are only known by legitimate users, and also it is necessary to clarify the possible values of those parameters. In the cryptosystem here considered the secret key is given by the eight initial conditions of the system and the coupling parameters. If those values are codified as in [Vidal *et al.*, 2012] using 11 digits, the cardinality of the key space is $(10^{11})^2 \times (10^{11})^8 = 10^{110}$ [4]. Nevertheless, the previous value is only correct if all parameters included in a secret key are independent and unknown by a potential attacker. This is not the case of the cryptosystem that we are analyzing. First of all, an eavesdropper has access to the values $x_{B_0}$ and $z_{A_0}$, since both values are transmitted in plain at the beginning of the communication session. Consequently, this implies a reduction of the key space and thus its cardinality is $(10^{11})^2 \times (10^{11})^6 = 10^{88}$.

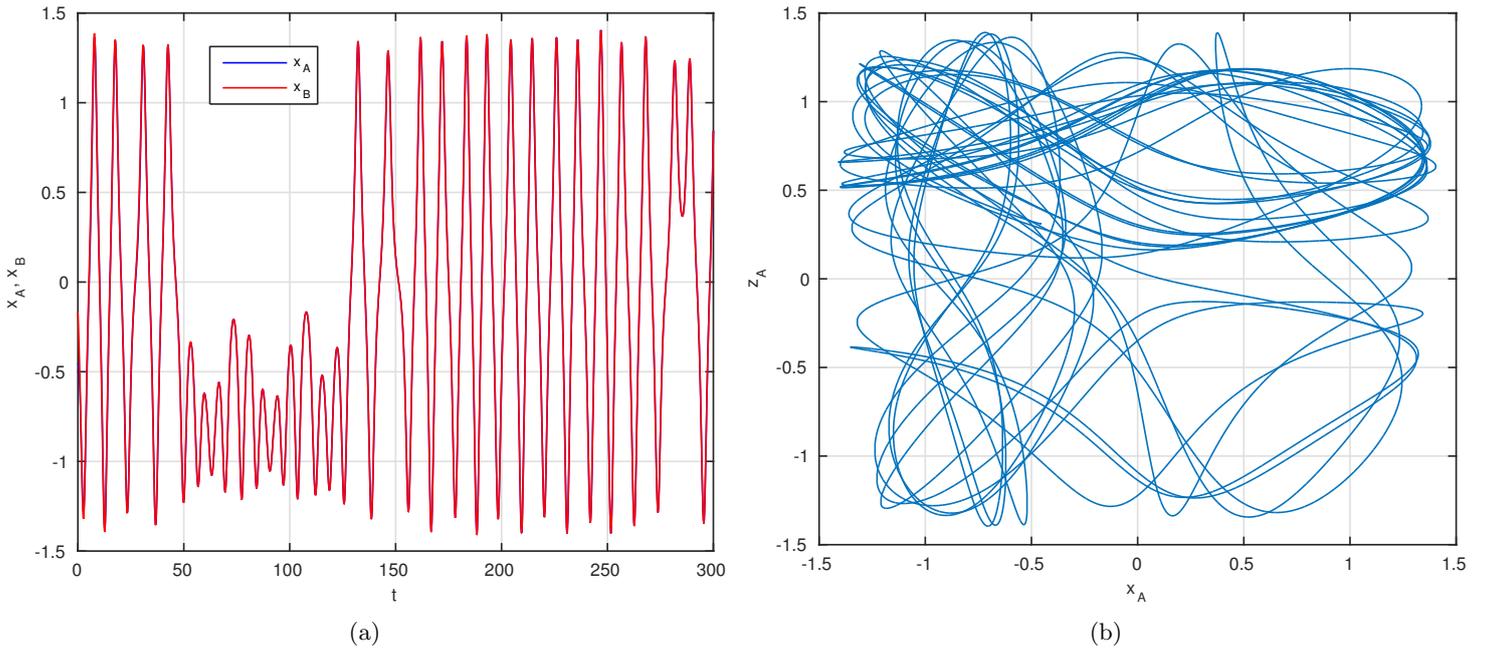

Fig. 4. (a) Trajectories of $x_A$ and $x_B$; (b) Projection of the attractor on the phase plane $(x, z)$. The configuration used in this simulation was: $a = -0.815215556019668$, $\mu = 0.697158139176817$, $b = 0.724394324457102$, $x_{A_0} = -0.45779369216014$, $y_{A_0} = -0.170731117605469$, $z_{A_0} = 0.312585918469052$, $w_{A_0} = -0.0302306179511633$, $x_{B_0} = -0.164151025323075$, $y_{B_0} = -0.324330970324339$, $z_{B_0} = -0.291053326006865$, $w_{B_0} = 0.405153559004464$.

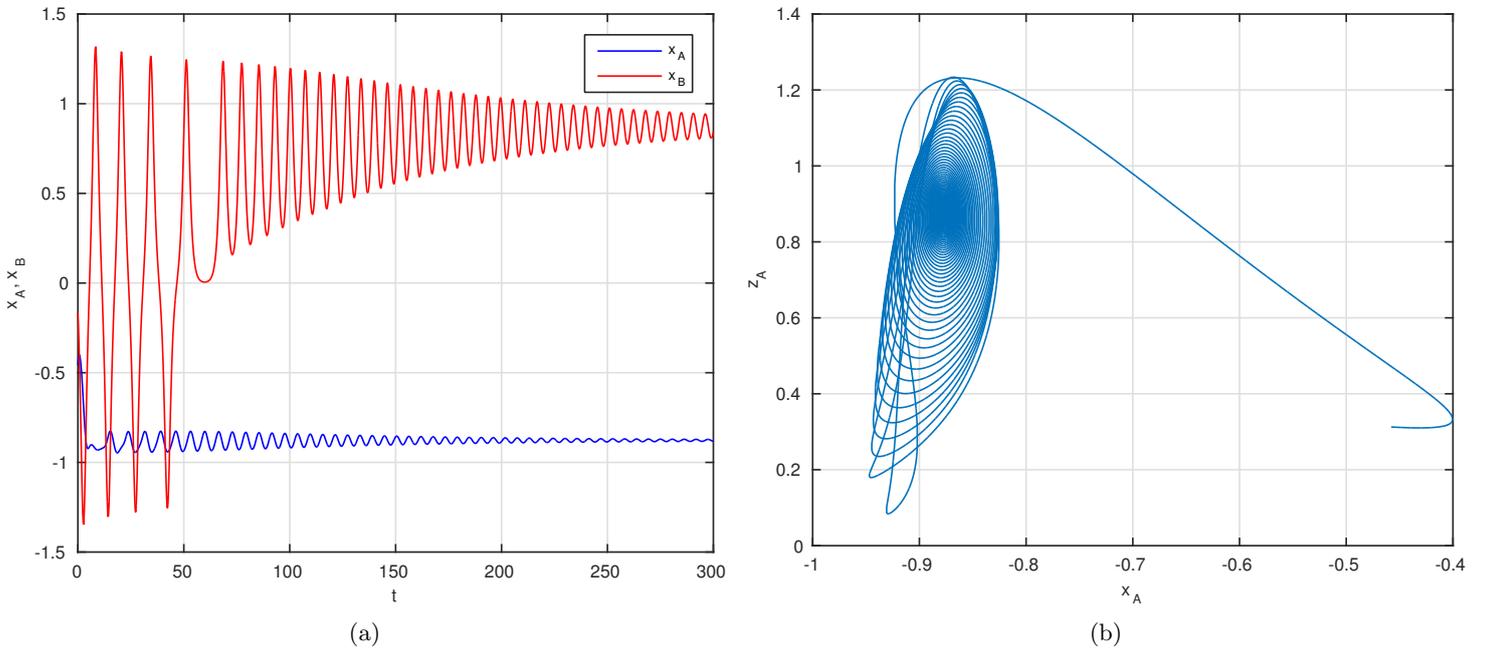

Fig. 5. Effect of a transmission delay of $10ms$ on synchronization: (a) trajectories of $x_A$ and $x_B$; (b) trajectory of the attractor projected on the phase plane $(x, z)$.

Secondly, an attacker does not need to get all initial conditions to recover the keystream. In fact, the attacker tipically possesses either the transmitter or the receiver and tries to infer the corresponding initial conditions and the coupling factor. Once the eavesdropper has recovered the initial conditions of Alice, or the ones of Bob, she can generate the keystream generated by the synchronization procedure. Hence, we have a further reduction of the key space which is of cardinality $10^{11} \times (10^{11})^3 = 10^{44}$.

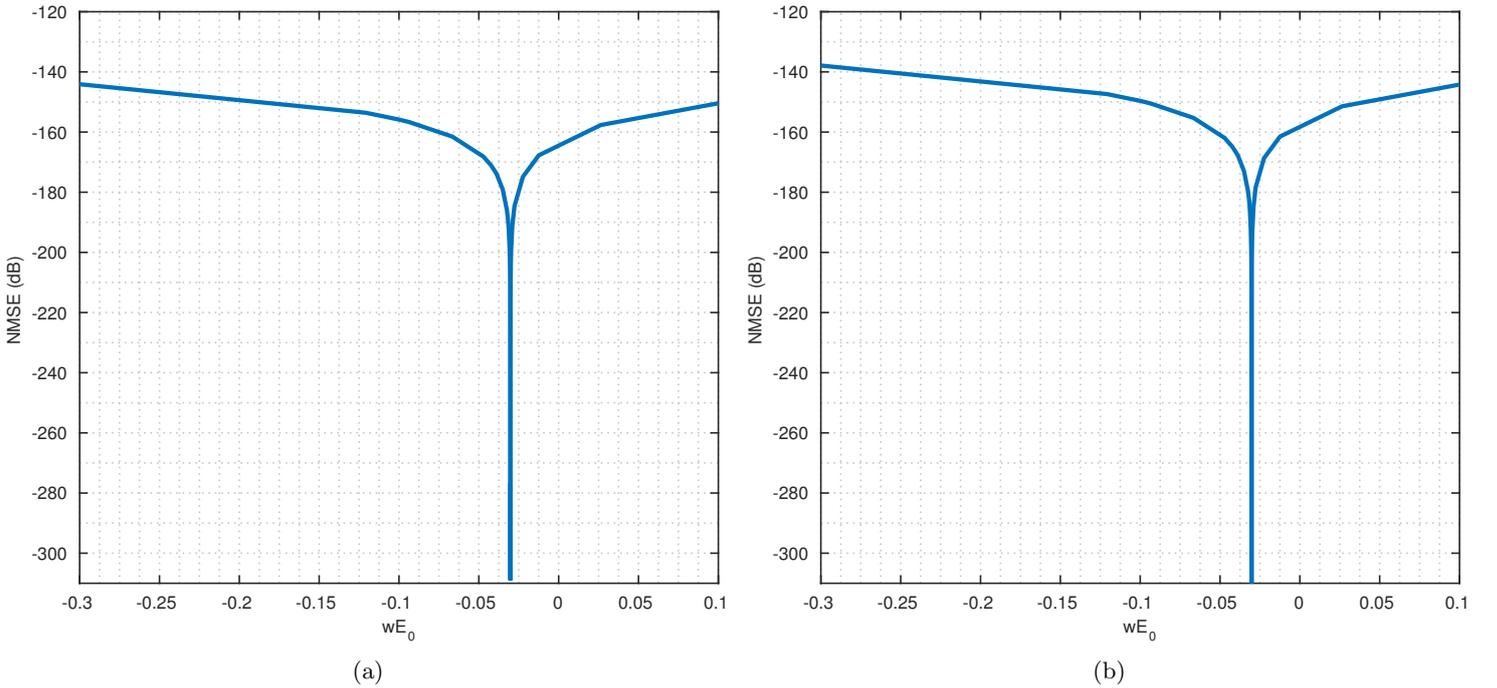

Fig. 6. NMSE associated to the estimation of $w_{E_0}$ from the only observable variable $z_A$. Two configurations were considered, sharing the same configuration for the control parameters ($a = -0.815215556019668$, $b = 0.724394324457102$, $\mu = 0.697158139176817$, $\epsilon_x = 0.797694334249407$, $\epsilon_z = 0.840527637336788$), along with the initial conditions for Alice ($x_{A_0} = -0.45779369216014$, $y_{A_0} = -0.170731117605469$, $z_{A_0} = 0.312585918469052$, $w_{A_0} = -0.0302306179511633$). The initial conditions for Bob were (a) $x_{B_0} = 0.289073514938958$, $y_{B_0} = 0.352263890343846$, $z_{B_0} = 0.00563661757175615$, $w_{B_0} = 0.135661388861377$; (b) $x_{B_0} = 0.238640291995402$, $y_{B_0} = 0.0859870358264758$, $z_{B_0} = -0.253265474014025$, $w_{B_0} = 0.166416217319468$.

### 4.2. *Information leaking through the analysis of the synchronism error*

On this point we are going to consider an attacker, Eve (E), that tries to reproduce Alice's system through the following set of equations:

---

[4]Here we should note that the definition of the key space in [Vidal et al., 2012] is not correct. The key space is determined by the set of all possible values for the parameters that are unkown by any potential attacker. In the cryptosystem under evaluation, these parameters are the eight initial conditions and the two coupling parameters.

$$\dot{x}_E = y_E + \varepsilon_{E_x}(x_B - x_E) \tag{9}$$
$$\dot{y}_E = \mu\, x_E + x_E \left(a(x_E^2 + z_E^2) + b z_E^2\right) \tag{10}$$
$$\dot{z}_E = w_E \tag{11}$$
$$\dot{w}_E = \mu\, z_E + z_E \left(a(x_E^2 + z_E^2) + b x_E^2\right) \tag{12}$$

Eve knows the value of $z_A$, since it is transmitted in clear by Alice. She knows also the value of $x_B$, which is sent in clear by Bob (take into account that Eve is actually performing a Man-In-The-Middle-Attack -MITMA-). To fully reproduce the system of Alice, Eve needs to determine exactly the values of the variables $x_{A_0}$, $y_{A_0}$, $w_{A_0}$, and the coupling factor $\epsilon_x$. Hereafter, we refer the estimation of such variables as $x_{E_0}$, $y_{E_0}$, $w_{E_0}$, and $\varepsilon_{E_x}$. From the general point of view of optimization theory, Eve tackles the inverse problem defined as

$$\arg\min_{\Theta} \frac{1}{T} \int_0^T \left(\frac{z_A - z_E}{z_A}\right)^2 dt \tag{13}$$

with $\Theta = (y_{E_0}, z_{E_0}, w_{E_0}, \varepsilon_{E_x})$, $z_A$ defined by Eq. (7) and $z_E$ determined through Eq. (11). At a first attempt to perform parameters estimation, Eve ponders the sensibility of the Normalized Mean Squared Error (NMSE) as implicitly defined in Eq. (13) with respect to the unknown control parameter and the initial conditions. As a matter of fact, $z_A$ is related to $w_E$ through a low-pass filtering, which can be further confirmed experimentally just by conducting a bi-search estimation of $w_{E_0}$ from the minimum value of the NMSE of $z_E$ with respect to $z_A$.

On these grounds, we have played the role of Eve performing 100 runs of the bi-search estimation using different configurations for Alice and Bob. In all the different configurations we have computed the Lyapunov exponents to confirm hyperchaoticity, and we have also verified that the synchronization is achieved. In the assumed MITMA scenario Eve does not know the value of $x_{A_0}$, $y_{A_0}$, $w_{A_0}$ and $\varepsilon_x$. Since Eve's goal in this stage is to get an estimation $w_{E_0}$ of $w_{A_0}$, in our experiments we have assigned random values to the other three unknown parameters. We have verified that for almost all the random values generated the function drawn by computing the NMSE with respect to $w_{E_0}$ is convex around the exact value of $w_{A_0}$. In Fig. 6 we can see this convexity for two settings of Alice. In fact, in our 100 experiments we have confirmed a mean error of order $10^{-10}$ in the estimation of $w_{A_0}$ through a naive bi-search algorithm. Nevertheless, it means a clear reduction of the sub-key space associated to such a variable. This being the case, the key space is (again) compressed to $10^{11} \times (10^{11})^2 \times 10^2 = 10^{35}$.

### 4.3. *Further reduction of the key space*

Recalling Eq. (13) for $\Theta = \varepsilon_x$, the next step in the security analysis is to determine whether is possible to get an estimation of other initial conditions. To achieve such a goal we are going to conduct a two-steps procedure. First, we perform a coarse grained exploration of the definition interval of $\varepsilon_x$, $x_{A_0}$, and $y_{A_0}$. In short, the definition space of $\varepsilon_x$ is splitted into $M$ equal-width intervals, and we keep the lower bounds of such intervals. The same procedure is applied to $x_{A_0}$ and $y_{A_0}$, although in this case the cardinality of the resulting set is $N$. As a result, we have got $M \times N^2$ possible Alice's configurations. The NMSE of $z_E$ with respect to $z_A$ is calculated, and we keep the values of $\varepsilon_{E_x}$, $x_{E_0}$ and $y_{E_0}$ that lead to a minimun value of the NMSE figure. Accordingly, we obtain an estimation of the unknown variables of Alice. In the second stage of our procedure we apply those estimations, along with the estimation of $w_{E_0}$, to perform an Ordinary-Differential-Equation (ODE) parameter fitting using some global optimization technique. In this work we have applied the pattern search algorithm for such a goal [Torczon, 1997], since the outcome of the previous stage informs about the existence of multiple local minima and that can be a problem when applying gradient-based optimization procedures. In order to confirm whether this two-fold methodology determine a further reduction of the key space, we have selected 100 different configurations leading to hyperchaotic behavior and the Matlab *patternsearch* function has been used for the parameters fitting. Along the different simulations it is possible to observe the convexity of the NMSE as defined in Eq. (13)

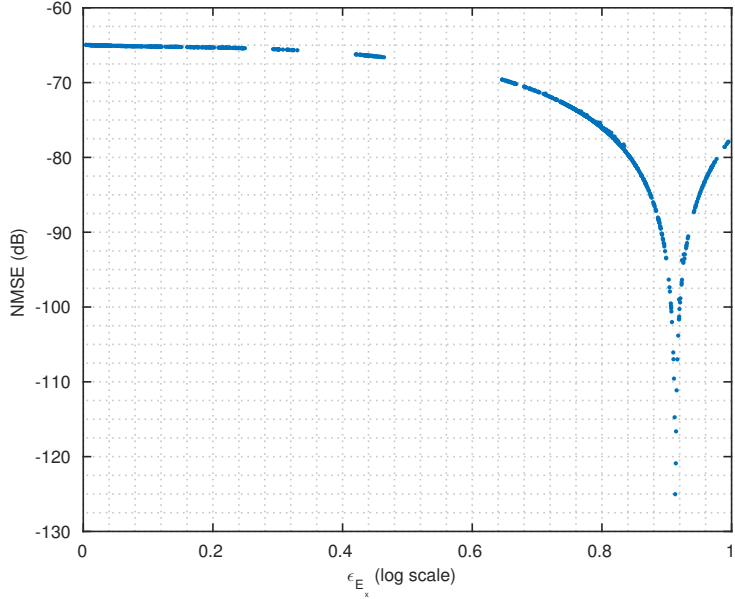

Fig. 7. NMSE with respect to $\varepsilon_{E_x}$ for $a = -0.905791937075619$, $b = 0.126986816293506$, $\mu = 0.814723686393179$, $\varepsilon_x = 0.913375856139019$, $\varepsilon_z = 0.63235924622541$, $x_{A_0} = -0.40245959500059$, $y_{A_0} = -0.221501781132952$, $z_{A_0} = 0.0468815192049838$, $w_{A_0} = 0.457506835434298$, $x_{B_0} = 0.00595705166514238$, $y_{B_0} = -0.244904884540731$, $z_{B_0} = 0.00595705166514238$, and $w_{B_0} = 0.199076722656686$.

Table 1. Distribution of the errors in the estimation of Alice's unknown secret parameters. Take into account that the error for $w_{A_0}$ is below $10^{-8}$ for the 17% of the considered configurations, smaller than $10^{-9}$ with a ratio of the 68%, and below of $10^{-11}$ for the 15% of the evaluated setups.

| Estimation error | $\approx 10^{-2}$ | $\approx 10^{-3}$ | $\approx 10^{-4}$ | $\approx 10^{-5}$ |
|---|---|---|---|---|
| $x_{A_0}$ | 1% | 52% | 43% | 4% |
| $y_{A_0}$ | 34% | 29% | 25% | 12% |
| $\varepsilon_x$ | 53% | 37% | 10% | 0% |

around the exact value of $\varepsilon_x$ (see Fig. 7). We have verified that in average the values of $x_{A_0}$ and $y_{A_0}$ can be obtained with an error below to $10^{-3}$, whereas $\varepsilon_x$ is estimated with a mean error below $10^{-2}$ (see Table 1).

Although the previous study highlights the reduction of the key space, a deeper examination makes possible to identify weaker selection of the keys. As a matter of fact, the selection of the control parameters according to [Vidal, 2011, p. 46] paves the way for an attacker to estimate the secret keys. This concern has been verified through 1000 experiments and considering that the initial conditions and the parameters are codified with 11 digits (i.e., in the vein of [Vidal et al., 2012]). Different random values were generated for the initial conditions and the coupling parameters. In each different configuration the gradient descend algorithm in [Orue et al., 2010] was used to get an estimation of $x_{A_0}$, $y_{A_0}$, $w_{A_0}$, and $\varepsilon_{E_x}$. As it is drawn from Fig. 8, a MITMA enables the complete recovery of the secret key in more than 25% of the considered setups. This is a major security problem that cannot be avoided due to the inner characteristics of the synchronization procedure used to encrypt information in [Vidal et al., 2012].

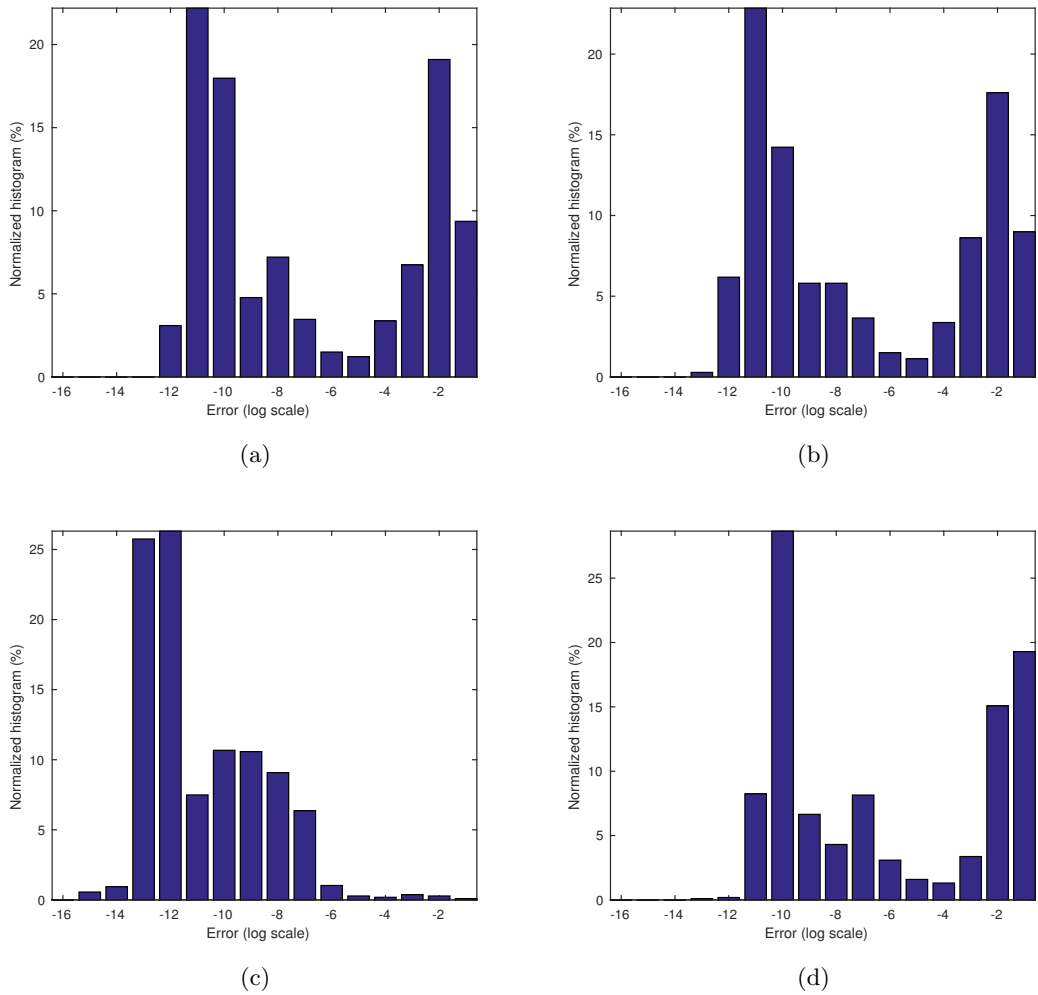

Fig. 8. Rate of estimation errors for 1000 setups with $a = -1$, $b = 0.9$, $\mu = 1.25$, and random values for the initial conditions and the coupling parameters. The plots show the normalized histogram for the estimation error of (a) $x_{A_0}$, (b) $y_{A_0}$, (c) $w_{A_0}$, and (d) $\varepsilon_x$.

## 5. Conclusion

In this paper we have highlighted some important weaknesses of the algorithm proposed in [Vidal *et al.*, 2012]. This cryptosystem shows serious efficiency problems, and it is possible to reduce drastically its key space just applying global optimization techniques. In fact, we have shown that in some cases it is possible to apply a MITMA to recover the secret key. All in all, our analysis shows that the cryptosystem defined in [Vidal *et al.*, 2012] does not provide a sufficient protection against brute force attacks given the computational power of today's computers.

In addition, the set of tools and the methodology applied in this paper can be very useful first to evaluate and perfect recent contributions in chaotic cryptography [Čelikovskỳ & Lynnyk, 2016; Shakiba *et al.*, 2016; Li, 2016; Xie *et al.*, 2017], and second to guide the design of new proposals. Regarding this last concern, it is necessary to recall that the computer implementation of chaos is affected by finite precision matters, which erases the link between chaos and cryptography as it was implicitly established by Shannon [Alvarez *et al.*, 2011]. If the encryption system is constructed by means of continuous-time chaotic systems, then we have to deal with the complexity associated to numerical integration methods and floating-point computation [Li, 2003; Abad *et al.*, 2012; Lozi & Pchelintsev, 2015]. Therefore, if chaos is the main bottom line of an encryption proposal, then one has to use electronic or optical devices to generate chaotic orbits and to lead chaotic communications and/or encryption [Lozi, 2014].

## Acknowledgments

This work was supported by Comunidad de Madrid (Spain) under the project S2013/ICE-3095-CM (CIBERDINE).

## References

Abad, A., Barrio, R., Blesa, F. & Rodríguez, M. [2012] "Tides, a taylor series integrator for differential equations," *ACM Transactions on Mathematical Software (TOMS)* **39**, 5.

Alvarez, G., Amigó, J. M., Arroyo, D. & Li, S. [2011] "Lessons learnt from the cryptanalysis of chaos-based ciphers," *Chaos-Based Cryptography*, eds. Kocarev, L. & Lian, S. (Springer Berlin Heidelberg), ISBN 978-3-642-20541-5, pp. 257–295, doi:10.1007/978-3-642-20542-2_8, URL http://dx.doi.org/10.1007/978-3-642-20542-2_8.

Arroyo, D., Alvarez, G., Amigó, J. M. & Li, S. [2011] "Cryptanalysis of a family of self-synchronizing chaotic stream ciphers," *Communications in Nonlinear Science and Numerical Simulation* **16**, 805–813, doi:http://dx.doi.org/10.1016/j.cnsns.2010.04.031.

Arroyo, D., Alvarez, G., Li, S., Li, C. & Fernandez, V. [2009a] "Cryptanalysis of a new chaotic cryptosystem based on ergodicity," *International Journal of Modern Physics B* **23**, 651–659, doi:http://dx.doi.org/10.1142/S0217979209049966.

Arroyo, D., Li, C., Li, S., Alvarez, G. & Halang, W. A. [2009b] "Cryptanalysis of an image encryption scheme based on a new total shuffling algorithm," *Chaos, Solitons and Fractals* **41**, 2613–2616, doi:http://dx.doi.org/10.1016/j.chaos.2008.09.051.

Baptista, M. S. [1998] "Cryptography with chaos," *Physics Letters A* **240**, 50–54.

Čelikovskỳ, S. & Lynnyk, V. [2016] "Message embedded chaotic masking synchronization scheme based on the generalized lorenz system and its security analysis," *International Journal of Bifurcation and Chaos* **26**, 1650140.

Chandre, C., Wiggins, S. & Uzer, T. [2003] "Time–frequency analysis of chaotic systems," *Physica D: Nonlinear Phenomena* **181**, 171–196.

Chen, G., Hsu, S.-B., Huang, Y. & Roque-Sol, M. A. [2011] "The spectrum of chaotic time series (ii): wavelet analysis," *International Journal of Bifurcation and Chaos* **21**, 1457–1467.

García-Martínez, M., Denisenko, N., Soto, D., Arroyo, D., Orue, A. & Fernandez, V. [2013] "High-speed free-space quantum key distribution system for urban daylight applications," *Applied Optics* **52**, 3311–3317, URL http://www.opticsinfobase.org/ao/upcoming_pdf.cfm?id=185412.

Hu, H., Deng, Y. & Liu, L. [2014] "Counteracting the dynamical degradation of digital chaos via hybrid control," *Communications in Nonlinear Science and Numerical Simulation* **19**, 1970 – 1984, doi:http://dx.doi.org/10.1016/j.cnsns.2013.10.031, URL http://www.sciencedirect.com/science/article/pii/S1007570413005200.

Li, C. [2016] "Cracking a hierarchical chaotic image encryption algorithm based on permutation," *Signal Processing* **118**, 203–210, doi:http://dx.doi.org/10.1016/j.sigpro.2015.07.008.

Li, C., Liu, Y., Xie, T. & Chen, M. Z. Q. [2013] "Breaking a novel image encryption scheme based on improved hyperchaotic sequences," *Nonlinear Dynamics* **73**, 2083–2089, doi:http://dx.doi.org/10.1007/s11071-013-0924-6.

Li, S. [2003] "Analyses and new designs of digital chaotic ciphers," PhD thesis, School of Electronic and Information Engineering, Xi'an Jiaotong University, Xi'an, China, available online at http://www.hooklee.com/pub.html.

Li, S., Alvarez, G., Li, Z. & Halang, W. [2007] "Analog chaos-based secure communications and cryptanalysis: a brief survey," *3rd Int. IEEE Scientific Conference on Physics and Control (PhysCon 2007)*, eds. Kurths, J., Fradkov, A. & Chen, G. (Potsdam, Germany), p. 92, full edition available at http://www.hooklee.com/Papers/PhysCon2007.pdf.

Liu, Y., Fan, H., Xie, E. Y., Cheng, G. & Li, C. [2015] "Deciphering an image cipher based on mixed transformed logistic maps," *International Journal of Bifurcation and Chaos* **25**, art. no. 1550188.

Lozi, R. [2014] "Designing chaotic mathematical circuits for solving practical problems," *International Journal of Automation and Computing* **11**, 588–597, doi:10.1007/s11633-014-0839-9, URL http://dx.

doi.org/10.1007/s11633-014-0839-9.

Lozi, R. & Pchelintsev, A. N. [2015] "A new reliable numerical method for computing chaotic solutions of dynamical systems: the chen attractor case," *International Journal of Bifurcation and Chaos* **25**, Article number 1550187.

Menezes, A., van Oorschot, P. & Vanstone, S. [1997] *Handbook of Applied Cryptography* (CRC Press).

Orue, A., Alvarez, G., Pastor, G., Romera, M., Montoya, F. & Li, S. [2010] "A new parameter determination method for some double-scroll chaotic systems and its applications to chaotic cryptanalysis," *Communications in Nonlinear Science and Numerical Simulation* **15**, 3471 – 3483, doi:10.1016/j.cnsns.2009.12.017, URL http://www.sciencedirect.com/science/article/pii/S1007570409006534.

Orue, A., Fernandez, V., Alvarez, G., Pastor, G., Romera, M., Montoya, F. & Li, S. [2009] "Breaking a SC-CNN-based Chaotic Masking Secure Communication System," *International Journal of Bifurcation and Chaos* **19**, 1329–1338, doi:10.1142/S0218127409023652, URL http://www.worldscientific.com/doi/abs/10.1142/S0218127409023652.

Shakiba, A., Hooshmandasl, M. R. & Meybodi, M. A. [2016] "Cryptanalysis of multiplicative coupled cryptosystems based on the chebyshev polynomials," *International Journal of Bifurcation and Chaos* **26**, Article number 1650112, doi:10.1142/S0218127416501121.

Shannon, C. [1949] "Communication theory of secrecy systems," *Bell Sys. Tech. J.* **28**, 656–715.

Sinai, Y. [1968] "Construction of Markov partitions." *Funct. Anal. Appl.* **2**, 245–253.

Torczon, V. [1997] "On the convergence of pattern search algorithms," *SIAM Journal on optimization* **7**, 1–25.

Vidal, G. [2011] "Sincronización y control de sistemas dinámicos en régimen de caos espacio-temporal," PhD thesis, Universidad de Navarra, Spain, URL http://dadun.unav.edu/handle/10171/17003.

Vidal, G., Baptista, M. S. & Mancini, H. [2012] "Fundamentals of a classical chaos-based cryptosystem with some quantum cryptography features," *International Journal of Bifurcation and Chaos* **22**, Article number 1250243.

Vidal, G. & Mancini, H. [2009] "Hyperchaotic synchronization under square symmetry," *International Journal of Bifurcation and Chaos* **19**, 719–726.

Wang, Q., Yu, S., Li, C., Lü, J., Fang, X., Guyeux, C. & Bahi, J. M. [2016] "Theoretical design and FPGA-based implementation of higher-dimensional digital chaotic systems," *IEEE Transactions on Circuits and Systems I-Regular Papers* **63**, 401–412.

Xie, E. Y., Li, C., Yu, S. & Lü, J. [2017] "On the cryptanalysis of fridrich's chaotic image encryption scheme," *Signal Processing* **132**, 150–154.